\def\bc{\begin{center}}                \def\ec{\end{center}}
\def\be{\begin{equation}}              \def\ee{\end{equation}}
\def\bear{\begin{eqnarray}}            \def\eear{\end{eqnarray}}
\def\la{\langle}      \def\ra{\rangle}     \def\l{\left}
\def\r{\right}        \def\dg{\dagger}      \def\ci{\cite}
\def\alf{\alpha}      \def\lb{\label}     \def\lam{\lambda}
\def\Lam{\Lambda}     \def\sig{\sigma}    \def\b{\beta}
\def\Dlt{\Delta}      \def\g{\gamma}        \def\o{\omega}
\def\t{\theta}        \def\vphi{\varphi}
     
\def\pr{\prime}       \def\td{\tilde}
                \def\rar{\rightarrow}
\def\lrar{\leftrightarrow}         \def\Lrar{\Leftrightarrow}
\def\sm{\small}
\def\bt{\begin{tabular}}          \def\et{\end{tabular}}

\documentstyle[12pt]{article}
\begin{document}

      \begin{flushright} Preprint quant-ph/9705001\\
                	  (Revised Aug 1997, Feb 1998)
      \end{flushright}
\bigskip

\begin{center}
    {\Large \bf On the squeezed states for {\large $n$} observables}
\end{center}
\medskip

\centerline     {{\large D.~A. Trifonov}\footnote[0]{$^+$e-mail:
				dtrif@inrne.acad.bg}$^{+}$}
\centerline     {Institute of Nuclear Research,}
\centerline     {72 Tzarigradsko Chauss\'ee,\,\,1784 Sofia, Bulgaria}
\vspace{8mm}

\begin{center}
\begin{minipage}{14cm}
\centerline{\sm {\bf Abstract}}
\vspace{5mm}

{\sm  Three basic properties (eigenstate, orbit and intelligence) of the
canonical squeezed states (SS) are extended to the case of arbitrary $n$
observables. The SS for $n$ observables $X_i$ can be constructed as
eigenstates of their linear complex combinations or as states which
minimize the Robertson uncertainty relation.  When $X_i$ close a Lie
algebra $L$ the generalized SS could also be introduced as orbit of
Aut$(L^C)$.  It is shown that for the nilpotent algebra $h_N$ the three
generalizations are equivalent. For the simple $su(1,1)$ the family of
eigenstates of $uK_- + vK_+$ ($K_\pm$ being lowering and raising operators)
is a family of ideal $K_1$-$K_2$ SS, but it cannot be represented as an
Aut$(su^C(1,1))$ orbit although the $SU(1,1)$ group related coherent
states (CS) with symmetry are contained in it.

Eigenstates $|z,u,v,w;k\ra$ of general combination $uK_- + vK_+ + wK_3$ of
the three generators $K_j$ of $SU(1,1)$ in the representations with
Bargman index $k=${\sm$1/2,1,\ldots$}, and $k=${\sm$1/4,3/4$} are
constructed and discussed in greater detail. These are ideal SS for
$K_{1,2,3}$.  In the case of the one mode realization of $su(1,1)$ the
nonclassical properties (sub-Poissonian statistics, quadrature squeezing)
of the generalized even CS $|z,u,v;+\ra$ are demonstrated. The states
$|z,u,v,w;k\!=\!\frac{1}{4},\frac 34\ra$ can exhibit strong both linear
and quadratic squeezing.}
\end{minipage}
\end{center}
\baselineskip=18pt \vspace{3mm}

\section{Introduction}

In the last decade or so a considerable attention was paid in the
literature to the squeezed states (SS), especially to the SS in quantum
optics \ci{LK}. In the one mode amplitude SS the variance of one of the two
quadratures $q$, $p$ of boson/photon annihilation operator $a$, $a =
(q+{\rm i}p)/${\sm$\sqrt{2}$}, can be reduced below its value of {\sm
$1/\sqrt{2}$} in the ground state $|0\ra$. We shall call these SS the
$q$-$p$ SS.  Most familiar one mode $q$-$p$ SS are the Stoler
$\zeta$-classes $|z,\zeta\ra$ \ci{SY}, the Yuen two photon coherent states
(CS) $|z,\mu,\nu\ra$ \ci{SY} and the Dodonov et. al.  correlated states
$|z,u,v\ra$ \ci{DKM}.  These three types of SS are equivalent \ci{T93} and
should be called one mode canonical SS (CSS) or standard SS. SS for other
pairs of observables are also considered in the literature
\ci{T94,WE,Vac}.  The canonical SS can be defined in the following three
equivalent ways \ci{T93} :\\[1mm]
a) as eigenstates of complex combination of $q$ and $p$:
	$(\b_1p+\b_2q)|{\rm CSS}\ra = z|{\rm CSS}\ra$ \ci{SY,MMT};
	\\[1mm]
b) as displaced and squeezed vacuum: $|{\rm CSS}\ra = S(\zeta)D(z)|0\ra$,
	where  $D(z)=\exp(za^\dg-z^*a)$, and $S(\zeta) =
	\exp[(\zeta a^{\dg 2}-\zeta^*a^2)/${\sm$2$}$]$ is the (canonical or
	ordinary) squeeze operator \ci{LK,SY};\\[1mm]
c) as states which minimize the Schr\"odinger inequality ($\Dlt X$ is the
variance of $X$ and $\Dlt XY$ is the covariance of $X$ and $Y$)
\be\lb{1}    
	\Dlt^2X\Dlt^2Y - \Dlt^2XY \geq \frac{1}{4}|\la[X,Y]\ra|^2 \ee
for $X=q$ and $Y=p$ \ci{DKM,T93}.
We note that the three equivalent definitions of one mode coherent states
(CS) $|z\ra$ \ci{Zh} are particular cases of the above three definitions
of CSS, namely $\b_1 = {\rm i}\b_2 = {\rm i}/\sqrt{2}$ in a), $\zeta=0$ in
b) and $\Dlt^2 qp = 0$ in c). Here $q$ and $p$ are dimensionless
quadratures of $a$.  Eigenstates of complex combination of {\it two}
observables $X$ and $Y$ are also called $X$-$Y$ SS or $X$-$Y$
Schr\"odinger intelligent states \ci{T94} (following \ci{DKM} they could
also be called Schr\"odinger correlated states). The term "intelligent
states" was introduced in \ci{Ar} on the example of spin states, which
minimize the Heisenberg inequality.

  The aim of this paper is to introduce SS for several observables $ X_i$,
$i=1,...,n$, and to construct and analyze SS for the Hermitian generators
$K_1,\,K_2,\,K_3$ of the group $SU(1,1)$.  The idea is to generalize the
above three basic properties of the CSS to the case of $n$ observables
$X_i$.
These possibilities stem from the observations that the product of the
canonical squeeze and displacement operators $S(\zeta)D(z)$ belongs
\ci{T96a} to the
group of automorphisms \ci{BR} of the Heisenberg-Weyl algebra $h_1$
(spanned by $p,\,q$,\, and the identity) and that the Schr\"odinger
inequality for {\it two} observables is a particular case of Robertson
uncertainty relation for $n$ observables \ci{Rob}.  

The paper is organized as follows. In section 2 we consider the possible
extensions of the definitions (a), (b) and (c) to the case on $n$ arbitrary
observables and discuss the problem of equivalence of generalized
definitions. The generalizations are based on the Robertson uncertainty
relation \ci{Rob} for $n$ operators and on the group of automorphisms
Aut$(L^C)$ of the corresponding complexified Lie algebra $L^C$ \ci{BR}.
We find that the first way of generalization (the eigenstate way) could be
considered as most general one. It is noted that not every continuous
family of eigenstates of combinations of $X_j$ can exhibit squeezing of
$X_j$. A sufficient condition for {\it ideal squeezing} (i.e., arbitrarily
strong squeezing) in such eigenstates is that of eq. (\ref{3}).

In section 3 we consider some examples of generalized SS for $n$
observables which are of current interest in physical literature,
especially in quantum optics:  the quadratures $p_\nu$, $q_\nu$ of $N$
boson destruction operators $a_\nu$, $\nu = 1,2,\ldots,N$ (generators of
the nilpotent Heisenberg-Weyl group $H_N$) and the quasi-spin components
$K_i$ (generators of the simple $SU(1,1)$). Here the eigenstates
$|z,u,v,w;k\ra$ of general complex combination of $K_j$ are constructed
explicitly (see also \ci{T96a,B97}) and shown to be ideal SS for the
generators $K_{1,2,3}$. The nonclassical properties of $|z,u,v,w;k\ra$ are
analyzed in the quadratic one mode bosonic representation.  It is
demonstrated that the generalized even CS $|z,u,v;+\ra$, which are
eigenstates of complex combination of $a^2$ and $a^{\dg 2}$ ($[ua^2/2 +
va^{\dg 2}/2]|z,u,v;\pm\ra = z|z,u,v;\pm\ra$) \ci{T95}, do exhibit
sub-Poissonian photon statistics (Fig. 2a) and strong linear and
quadratic quadrature squeezing (Fig. 1a) [The variance of a quadrature
component of $a$ or $a^2$ is said to be squeezed if it is less than its
value in the ground state $|0\ra$].  Moreover, there are states from this
subfamily, which can exhibit linear and quadratic squeezing simultaneously
(joint squeezing).  These joint SS, and all $|z,u,v;\pm\ra$ as well, can
be generated using the scheme, which is a modification of the recently
proposed scheme of Brif and Mann \ci{BM97}.  

We note that the first example of generalized SS (section 3) for the $n$
canonical operators $p_\nu,\,\,q_\nu$ is most symmetric: as in the one
mode case, here the three generalized definitions are equivalent.
The second example of generalized SS (for the generators of $SU(1,1)$)
does not possess such symmetry:
examples of continuous families  of eigenstates of $su(1,1)$ operators
(generally called $su(1,1)$ algebraic CS \ci{T96a}) are pointed out which
are not Aut$(su^C(1,1))$ orbits for any reference state $|\psi_0\ra$.
Such are the sets of the $K_1$-$K_2$ SS $|z,u,v;k\ra$ (eigenstates of
$uK_- + vK_+$) \ci{T94}, the Barut-Girardello CS (eigenstates of $K_-$)
\ci{BG} and the even and odd CS (eigenstates of $a^2$) \ci{DMM}. 

\section{Generalized SS}
\subsection{Generalization of the eigenvalue property}

The generalization of property a) is straightforward:  We introduce the
shortened notation $\vec{X} = (X_1,X_2,\ldots,X_n)$ and consider the sets
of eigenstates of complex combinations  of all $X_i$ (summation over
repeated indices is adopted),

\be\lb{2}  
A_\nu(\b)|\vec{z},\b\ra = z_\nu|\vec{z},\b\ra,\quad
A_\nu(\b)=\b_{\nu i}X_i,
\ee
where $\b$ is an $n_{\rm c}\times n$ complex matrix and the integer
$n_{\rm c}$ is to be yet specified.  The greek indices $\mu,\,\nu$ run
from $1$ to $n_{\rm c}$, and the latin indices $i,\,j$ run from $1$ to $n$.
Eigenstates $|\vec{z},\b\ra$ would exhibit arbitrarily strong squeezing
({\it ideal squeezing}) of the observable $X_j$ when for a given $\nu$ all
but $\b_{\nu j}$ are let to tend to $0$, \be\lb{3}    
\longrightarrow 0 \quad {\rm for\,\, all}\quad k\neq j\quad {\rm and\,\,
at\,\, least\,\, one}\quad \nu.  \ee This  general possibility (for one
  mode CSS it is easily verified) stems from the observation \ci{T94},
that the variance $\Dlt X$ of an Hermitian operator $X$ vanishes in pure
states $|\psi\ra$ if and only if $|\psi\ra$ is an eigenstate of $X$:
\be\lb{4}   
X|\psi\ra = x|\psi\ra \quad \Lrar \quad \Dlt X(\psi) = 0.
\ee
In the limit (\ref{3}) the state $|\vec{z},\b\ra$ tends to an
eigenstate of $X_j$ and this ensure arbitrarily strong squeezing of
$X_j$, i.e. $\Dlt X_j \rar 0$.  To observe light squeezing one needs not
to take the limit in (\ref{3}).  The variance $\Dlt X$ may vanish in
mixed states $\rho = \sum p_n|\psi_n\ra\la\psi_n|$ if all $|\psi_n\ra$ are
eigenstates of $X$ with the same eigenvalue.

{}For several observables we need a definition of the {\it family of SS}:
A family of states $|\psi(l_1,l_2,\ldots)\ra$ with parameters
$l_1,l_2,\ldots$ is  called a {\it family of SS for $n$ observables} $X_j$
(shortly $\vec{X}$-SS) if for every $j=1,2,...n$ one can find in it
states, such that $\Dlt X_j$ is less than a certain value $\Dlt_0$,

\be\lb{5}    
\Dlt X_j < \Dlt_0, \quad j = 1,2,\ldots,n\,.
\ee
The reference value $\Dlt_0 > 0$ is the variance of some of $X_i$ in some
reference state $\psi_0$, selected on certain physical reason, $\Dlt_0 =
\Dlt X_i(\psi_0)$.   One natural criterion  for $|\psi_0\ra$ is to
provide the equality of two or more $\Dlt X_j$ on the as lowest possible
level,

\be\lb{Dlt_0}   
\Dlt_0 =
{\rm Min}\{\Dlt X_1(\psi),\Dlt X_2(\psi),\ldots \}
\quad{\rm provided} \quad
\Dlt X_1(\psi) =\Dlt X_2(\psi) =\ldots .
\ee
Note that we work here with dimensionless operators $X_j$. For the
electromagnetic field (in quantum optics) one usually takes $|\psi_0\ra
=|0\ra$, $|0\ra$ being the vacuum.  For the quadratures of any
power of the annihilation operators $a_\nu$, $\nu = 1,2,\ldots,N$, the
choice $|\psi_0\ra =|0\ra$ ensures the equalities $\Dlt X_1(\psi_0) =\Dlt
X_2(\psi_0) =\ldots = \Dlt X_n(\psi_0) \equiv
\Dlt_0$ on the lowest level.

In the family $\{|\vec{z},\b\ra\}$ the SS defining inequality (\ref{5}) is
universally (i.e. for any choice of $\Dlt_0 > 0$) satisfied if conditions
(\ref{3}) hold. A set of states in which the inequality (\ref{5}) can hold
for an arbitrarily small $\Dlt_0$ (not simultaneously), i.e.,

\be\lb{6}    
\Dlt X_j \rar 0, \quad j=1,2,\ldots, n \,\,,
\ee
should be called a set of {\it ideal $\vec{X}$-SS}. Conditions (\ref{3})
ensure  (\ref{6}). The known CSS constitute such ideal SS for $q$ and $p$.

It is worth noting that the ideal squeezing conditions (\ref{3}) (and even
the conditions (\ref{5})) require enough parameter freedom in the family
$\{|\psi(l_1,l_2,\ldots)\ra\}$.  In the case of eigenstates
$|\vec{z},\b\ra$ of $A_\nu(\b)$ this means that parameters $\b_{\nu i}$
should not be fixed, i.e., $|\vec{z},\b\ra$ should be eigenstates of a set
of complex combinations of $X_j$. The number of free parameters
$\b_{\nu i}$ should evidently be not less than the number $n$ of the
observables.  Then from this set on can form several (say $n_{\rm c}$) linearly
independent operators $A_\nu(\b)$.

It is desirable to have the possibility to calculate in $\vec{X}$-SS all
second moments of $X_j$ in pure algebraic way as it is the case of CSS.
For even $n$ we can perform this for states (\ref{2}) if $n_{\rm c} = n/2$.
In order to do this we denote $n/2 = N$ and introduce the $n$ component
vector

$$\vec{B}(\b) =
(A_1(\b),\ldots,A_{N}(\b),A^\dg_{1}(\b),\ldots,A^\dg_{N}(\b)),\quad A_\nu
= \b_{\nu i}X_i). $$
Next we express $\vec{X}$ in terms of  $\vec{B}(\b)$. Then after some
calculations we obtain for the uncertainty matrix $\sig$ (an $n\times n$
matrix with elements
$\sig_{ij} = \la X_iX_j+X_jX_i\ra/2 - \la X_i\ra\la X_j\ra$)
in SS $|\vec{z},\b\ra$ the
following general expression in terms of first moments of commutators
$[X_i,X_j]$,

\be\lb{8}  
\sig({\vec{X}};\b) = {\cal B}^{-1}\l( \bt{cc} $0$ & $C^\pr$ \\
$C^\pr{}^{\rm T}$ & $0$ \et \r){\cal B}^{-1}{}^{\rm T},\quad{\cal B} = \l(\bt{cc}
$\b^{(1)}$&$\b^{(2)}$\\  $\b^{(1)}{}^*$&$\b^{(2)}{}^*$\et\r),
\ee
where $C^\pr = (C^\pr_{\nu\mu})$,
$$C^\pr_{\nu\mu} = \la[A_\nu(\b),A^\dg_\mu(\b)]\ra/2 =
\b_{\nu i}\b^*_{\mu j}\,[X_i,X_j],$$
$C^{\rm T}$ is the transposed $C$ and the $N\times N$ matrices $\b^{(1,2)}$
are defined as $\b^{(1)}_{\nu\mu} = \b_{\nu\mu}$, $\b^{(2)}_{\nu\mu}
= \b_{\nu,N+\mu}$. Note that $\b$ in (\ref{2}) now ($n_{\rm c}=N$) is
an $N\times n$ matrix, while ${\cal B}$ in (\ref{8}) is $n\times n$. We
suppose that ${\cal B}$ is not singular.  This is equivalent to the
nonsingularity of the linear real transformation of the observables

\be\lb{9}    
\vec{X} \rar \vec{X^\pr}= \Lam\vec{X},\quad \vec{X^\pr}=
(X^\pr_1,\ldots,X^\pr_n),
\ee
where $X^\pr_{\nu}$ and $X^\pr_{N+\nu}$ are quadrature components of
$A_\nu(\b)$, $A_\nu(\b)= X^\pr_\nu +iX^{\pr}_{N+\nu}$.  The $n\times n$
real matrix $\Lam$ is simply composed in terms of matrix elements of
$\b$:  $\Lam_{\nu i} = {\rm Re}\,\b_{\nu i}$, $\Lam_{N+\nu, i} = -{\rm
Im}\,\b_{\nu i}$.  Formula (\ref{8}) could be extended to the case of odd
$n$ if $n_{\rm c} = [n/2]$ ( $[n/2]$ is the integer part of $n$) and we
admit that $|\vec{z},\b\ra$ are eigenvectors of one extra Hermitian
operator $X^\pr_{2n_{\rm c}+1}$.

Thus $n_{\rm c}=[n/2]$ eigenvalue equations (\ref{2}) and the
nonsingularity of the transformation (\ref{9}) provide a proper
generalization of the first definition a) of the CSS to the case of $n$
arbitrary observables. The family of eigenstates $|\vec{z},\b\ra$ can be
qualified as a family of strong (or ideal) SS for $n$ operators $X_j$ if
the parameters $\b_{\nu j}$ could obey the conditions (\ref{5}) (or
(\ref{3})). Some squeezing is not excluded in other parameter ranges. We
will below see that, besides the useful formula (\ref{8}), the requirement
$n_{\rm c} = [n/2]$ in (\ref{2}) and (\ref{9}) provide an efficient
generalization of the third property c) of one mode CSS as well.

\subsection{Generalization of the orbit property of CSS}

When $X_i$ close a Lie algebra $L$ the second property of CSS can
be generalized in the form

\be\lb{10}   
|\psi_{\rm GSS}(g)\ra = U_{\rm A}(g)|\psi_0\ra,
\ee
where $U_{\rm A}(g)$ is an unitary representation of the group $G_{\rm A}
\equiv {\rm Aut}(L^C)$ of {\it automorphisms} of $L^C$ ($L^C$ is the
complexified $L$) \ci{BR} and $|\psi_0\ra$ is eigenvector of a fixed
element $A_0$ of $L^C$,

\be\lb{11}    
A_0|\psi_0\ra = z_0|\psi_0\ra,\quad A_0 \in L^C.
\ee
The idea of the definition (\ref{10}) of SS for $n$ Lie group generators
$X_j$ came from the observation \ci{T93} that the product $D(z)S(\zeta)$
of the displacement $D(z)$ and squeezed $S(\zeta)$ operators, which
appears in the second definition b) of CSS, is an element of the
semidirect product group $SU(1,1)\!\subset\!\!\!\!\!\!\times H_1$ of the
quasi-unitary group $SU(1,1)$ and the Heisenberg-Weyl group $H_1$.  And
$SU(1,1)\!\subset\!\!\!\!\!\!\times H_1$ is the group of automorphisms
Aut$(h^C_1)$ of the complexified algebra $h^C_1$, spanned by the canonical
observables $q$, $p$ and the identity.

Not all of the states $|\psi_{\rm GSS}(g)\ra$ however can exhibit
squeezing in the Hermitian operators $X_i \in L$. The standard $G$-group
related CS with symmetry $|\psi(g)\ra$ \ci{Zh} constitute such exceptions
($G$ being the group generated by $L$).  Indeed,  it is clear that
$|\psi(g)\ra$   are of the form of $|\psi_{\rm GSS}(g)\ra$, eq.
(\ref{10}), since Aut$(L^C)$ contains $G$.  More precisely, $G$ is
homomorphic to the group Ad$(L)$ of internal automorphisms of $L$ which is
a subgroup of Aut$(L^C)$:

$${\rm Ad}(L) \subset {\rm Ad}(L^C)\subset {\rm Aut}(L^C).$$

The $G$-group related CS with symmetry are eigenstates of
$U(g)A_0U^{-1}(g)$ which is a very particular combination of the
generators $X_j$: if $A_0$ is one of $X_j$, say $X_k$, then condition
(\ref{3}) can holds for $j=k$ only; even if $|\psi_0\ra$ is eigenvector of
several $A_0$ neither the inequalities (\ref{5}) for $|\psi_0\ra$ nor the
conditions (\ref{3}) could be satisfied for all $j$.
The absence of squeezing (in the sense of definition (\ref{5}) with
(\ref{Dlt_0})  in spin ($SU(2)$) and quasi-spin ($SU(1,1)$) group related
CS with symmetry was noted in \ci{T94}.  Thus one has to look for
squeezing of elements $X_j$ of $L$ in the states of orbits of the larger
group Aut$(L^C)$ of automorphisms of $L^C$, not in the states of orbits
of $G$. However for semisimple Lie algebras even this extension of the
group related CS may be insufficient to include ideal SS as we shall see
on the example of $su(1,1)$.

Let us examine briefly the relationship between two generalizations
(\ref{2})  and  (\ref{10}). By the definition of the group of
automorphisms $G_{\rm A}$ the operator $U_{\rm A}(g)A_0U^{-1}_{\rm A}(g)$
is a complex combination of $X_j$.  Noting that $|\psi_{\rm GSS}(g)\ra$ is
an eigenstate of $U_{\rm A}(g)A_0U^{-1}_{\rm A}(g)$ (with eigenvalue
$z_0$) we obtain that (\ref{10}) is equivalent to (is reduced to)
(\ref{2}).  The inverse however is generally not true:  eigenstates
$|\vec{z},\b\ra$ of some combinations $\b_jX_j$ may not be represented in
the form (\ref{10}) with unitary $U_{\rm A}(g(\b))$ and fixed, $\b$ and
$g$ independent reference vector $|\psi_0\ra$ (i.e. in the form of
Aut$(L^C)$-group related CS). The standard even and odd CS $|\alf\ra_\pm$
\ci{DMM}, the Barut-Girardello CS (BG CS) $|z;k\ra$ \ci{BG} and their
generalizations $|z,u,v;k\ra$ \ci{T94} are  examples of such families
which are neither $G$ nor Aut$(L^C)$ group related CS (proof in Appendix
A).  At the same time the states $|z,u,v;k\ra$ are ideal SS for $K_1$ or
$K_2$ \ci{T94} since they do satisfy the conditions (\ref{3}) and can
exhibit arbitrarily strong squeezing of $K_1$ or $K_2$ \ci{T94,T96a,T96b}.
This motivates the necessity to introduce the more general notion of {\it
$L$-algebraic CS} \ci{T96a} to denote a continuous family of states, which
are eigenstates of operators of the complexified $L^C$.  Thus the set of
$SU(1,1)$-group related CS with symmetry and the three sets of
$|\alf\ra_\pm$, $|z;k\ra$, and $|z,u,v;k\ra$ are all proper
$su(1,1)$-algebraic CS.   In the next section we construct most general
$su(1,1)$-algebraic CS and show that they are ideal $su(1,1)$ SS.

\subsection{Generalization of the intelligence property of CSS}

The third basic property of one mode CSS can be generalized using the
Robertson uncertainty relation (RUR) for $n$ Hermitian operators $X_i$
\ci{Rob},
\be\lb{12}
\det\sig(\vec{X}) \ge \det C(\vec{X}),\quad C=(C_{kj}),\,\,
C_{kj} = (-{\rm i}/2)\la [X_k,X_j]\ra,
\ee
where $\sig$ is the uncertainty matrix, defined in the subsection 2.1. The
inequality (\ref{12}) is valid for any state, pure or mixed.  For two
operators it coincides with the Schr\"odinger relation, eq. (\ref{1}). The
third way of construction of SS for $n$ observables could be as states
$|\psi\ra$, which minimize the RUR,

\be\lb{13}  
\det\sig(\vec{X};\psi) = \det C(\vec{X};\psi).
\ee
The properties of the uncertainty matrix $\sig(\vec{X})$  for $n$
observables and the minimization of RUR are studied in detail in ref.
\ci{T97}.  It is proven that for any $n$ the RUR is minimized in a state
$|\psi\ra$ if $|\psi\ra$ is an eigenstate of a real combination $\b_iX_i$.
For odd $n$ this condition is also necessary. For even $n$ the RUR is
minimized in eigenstates $|\vec{z},\b\ra$ of $n/2$ complex combinations of
$X_i$ (see eq.  (\ref{2}) with $n_{\rm c} = n/2$).  By direct calculation one
can check that the uncertainty matrix $\sig(\vec{X},\b)$, eq. (\ref{8}),
satisfies the equality (\ref{13}).

In addition to the results of ref. \ci{T97} we note here that in case of
{\it two} arbitrary observables $X$ and $Y$ the eq.  (\ref{2}) (with
$n_{\rm c}=1$) is also a necessary condition for a pure state to minimize
RUR.  For this purpose consider the mean value of nonnegative operators
$F^\dg(\b,r)F(\b,r)$, where $F = \b X +{\rm i}rY-(\b\la X\ra+{\rm i}r\la
Y\ra)$, $\b$ is complex and $r$ is a real parameter. This mean value is
easily expressed as a linear combination of three second moments of $X$
and $Y$ and the first moment of their commutator.  Then after some simple
algebra one gets the result that if a mixed state $\rho =
\sum_m\rho_m|\psi_m\ra\la\psi_m|$ minimizes the RUR for $n=2$ then
$F(\b,r)|\psi_m\ra=0$ for every $m$.  We anticipate that for any even $n$
the eqs. (\ref{2}) with $n_{\rm c}=n/2$ are again necessary.

States which minimize (\ref{12}) should be called {\it Robertson
intelligent states} (RIS), and in case of $n=2$ they should be referred to
as Schr\"odinger IS. If the set $\{X_i\}$ closes an algebra $L$ then the
minimizing states should be also referred to as $L$-algebra RIS.  It was
proven \ci{T97} that group related CS with symmetry for semisimple Lie
groups are RIS for group generators and CS with maximal symmetry are RIS
also for the quadratures of Weyl lowering operators. Thus group related CS
with symmetry are subset of the corresponding $L$-algebra RIS.

When the covariances (the correlations) of $X_i,\,X_j$ are not vanishing
the minimizing states should be called also Robertson correlated states,
following the ref.  \ci{DKM}. It was proven however that in any state the
correlations can be canceled by means of linear orthogonal or symplectic
transformations of the observables \ci{T97}.

RIS could exhibit arbitrarily strong squeezing of the observables if the
conditions (\ref{3}) can be satisfied. Normally this is the case:  RIS
generally depend on complex parameters $\b_{\nu i}$ and $z_\nu$.  The
total number of real parameters (for even $n$) is equal to $3n^2/2$.  If
$A_\nu(\b)$ are not subjected to further constrains we have enough
parameter freedom to ensure (\ref{3}). Squeezing may occur also in the
cases of eigenstates of $A_\nu(\b)$ with $n_{\rm c} < [n/2]$ which could not
minimize the Robertson relation for all $n$ observables. Note that the
large set of RIS contains many well defined subsets, states of which
cannot exhibit squeezing - such are the group related CS with symmetry for
semisimple (and some nonsemisimple) Lie groups.

The Robertson uncertainty relation provides a natural specification of the
reference value $\Dlt_0$ and the reference state $|\psi_0\ra$ in the
definition (\ref{5}) and (\ref{Dlt_0}) of SS for $n$ observables. For even
$n$, $n=2N$, one can require

\be\lb{14}    
\Dlt X_j(\psi) < \Dlt_0 = {\rm
Min}\l[\det C(\vec{X};\psi_{\vec{z}})\r]^{1/n},
\ee
provided
\be\lb{15}    
(X_\nu+{\rm i}X_{N+\nu})|\psi_{\vec{z}}\ra = z_\nu|\psi_{\vec{z}}\ra,
\quad \nu = 1,\ldots,N , \ee the minimization being with respect to
$z_\nu$.  The eigenvalue equations (\ref{15}) ensure the equality in
Robertson relation with equal variances $\Dlt^2 X_\nu(\vec{z}) = \Dlt^2
X_{N+\nu}(\vec{z}) = |\la\vec{z}|[X_\nu,X_{N+\nu}]|\vec{z}\ra|/2$.

The Eberly-Wodkiewicz definition of relative squeezing \ci{WE} can be
extended as follows: The variances $\Dlt X_j(\psi)$ are squeezed if

\be\lb{16}   
\Dlt X_j(\psi) < \l[\det C(\vec{X};\psi)\r]^{1/n},
\ee
This definition fails in the case of $X_j$ with discrete spectrum: in the
eigenstates of $X_j$ one has absolute squeezing, $\Dlt X_j = 0$, and $\det
C(\vec{X};\psi)\ra = 0$. For odd number $n$ both inequalities (\ref{14})
and (\ref{16}) fail since $\det C(\vec{X};\psi) = 0$ in any state. In
these cases one can apply the universal definitions (\ref{3}) and
(\ref{5}) for ideal and/or strong SS.  We shall demonstrate this in the
next section on the example of three generators of the $SU(1,1)$ group.
\vspace{5mm}

\section{Examples of SS and RIS for several observables}

In this section we consider two explicit constructions of generalized SS
and RIS. The first one is related to the nilpotent Lie algebra $h_N$ which
has $2N+1$ basic elements ($2N$ observables) and the second one is related
to the simple $su(1,1)$, which has $3$ basic elements ($3$  observables).
These algebras and observables are most frequently used in physics,
especially in quantum optics.

\subsection{$h_N$ SS and RIS}

The Heisenberg-Weyl algebra $h_N$ is spanned by $p_\mu,\,\,q_\mu$
($\mu=1,\ldots,N$) and $1$.  $p_\mu$ and $q_\mu$ are quadrature components
of boson operators $a_\mu$, $a_\mu = (q_\mu +
ip_\mu)/\sqrt{2}$. Here $N$ independent and mutually commuting linear
combinations $A_\mu(\b)$ of $a_\nu$, $a^\dg_\nu$ (equivalently of
$q_\nu$ and $p_\nu$) always exist, so that $n_{\rm c}=N$. The solution of eqs.
(\ref{2}) for operators $\vec{A}(\b) = (A_1(\b), A_2(\b),\ldots,
A_N(\b))$

\be\lb{17a}   
\vec{A}(\b) = \b^{(1)}\vec{p} + \b^{(2)}\vec{q}
\ee
exists for nonsingular $\b^{(1)}$ and arbitrary $\b^{(2)}$. In
coordinate representation it takes the form of an exponential of a
quadratic,

\be\lb{17b}   
\la\vec{q}|\vec{z},\b\ra = \td{\cal N}\exp[\vec{q}{\cal M}\vec{q} +
\vec{\cal N}\vec{q}],\,\quad {\cal M} = -({\rm i}/2){\b^{(1)}}^{-1}
\b^{(2)}, \,\, \vec{\cal N}={\rm i}{\b^{(1)}}^{-1}\vec{z},
\ee
We see a freedom in the set of parameters $\b_{\nu i}$ which can always be
used to subject $A_\nu(\b)$ to the canonical commutation relations, %

\be\lb{18}       
[A_\nu(\b),A^\dg_\mu(\b)] = \delta_{\nu\mu}.
\ee
Eigenstates of $N$ boson operators $A_\nu$, which are linear combinations
of $q_\nu$, $p_\nu$ (therefore of $a_\nu$ and $a^\dg_\nu$) and satisfy
(\ref{18}), were first constructed in \ci{MMT}.

The uncertainty matrix in states $\la\vec{q}|\vec{z},\b\ra$ is given by
the general formula (\ref{8}) with $C^\pr = (${\sm$1/2$}$)1_N$ and one
can check that it satisfies the equality in (\ref{12}). Therefore these
solutions are $h_N$ RIS. Moreover, even if a state $|\psi\ra$ is
an eigenstate of one complex combination of $p_\nu$ and $q_\nu$, then there
exist $N$ independent such combinations which have this $|\psi\ra$ as
their common eigenstate, i.e. $|\psi\ra$ is RIS and takes the form (\ref{10})
(proof in Appendix A).

The inverse is also true. If a state $|\psi\ra$ minimizes the Robertson
inequality for $p_\nu$ and $q_\nu$ then $|\psi\ra$ is an eigenstate of $N$
complex combinations $A_\mu$ of $p_\nu$ and $q_\nu$ in the form of new
boson operators (proof in \ci{T97,T95}). In
view of Aut$(h^C_N) = Mp(N,R)\!\subset\!\!\!\!\!\!\times H_N$ ($Mp(N,R) =
\overline{Sp(N,R)}$) we get that $N$ mode canonical RIS are of the form of
(\ref{10}) with $G_{\rm A} = Mp(N,R)\!\subset\!\!\!\!\!\!\times H_N$.

In quantum optics eigenstates of complex combinations of $p_\nu,\,q_\mu$
(i.e. the $h_N$ RIS, given in coordinate representation by (\ref{17b})) are
known as  (canonical) {\it multimode} SS and their nonclassical properties
have been intensively studied (see \ci{Ma} and references therein). The
$h_N$ RIS (\ref{17b})) obey the conditions (\ref{3}) and can exhibit
arbitrarily strong squeezing in $p_\nu$ or $q_\nu$, i.e., they are ideal
SS. The limits in (\ref{3}) however can not be taken "till the end", since
$p_\nu$ and $q_\nu$ have no normalizable eigenstates and $\Dlt q_\nu >
0$, $\Dlt p_\nu > 0$. Moreover no real combination of $p_\nu$ and
$q_\nu$ can be diagonalized \ci{T97}.

Thus we have shown that for canonical observables $p_\nu,\,q_\nu$ of $N$
mode boson system, i.e., for $h_N$ algebra, all three definitions a), b)
and c) of the one mode CSS are equivalently generalized on the basis of
the Robertson uncertainty relation and the group of automorphisms of
$h_N$.  We shall see below that such equivalence does not occur in the
case of $su(1,1)$ algebra observables.

\subsection{$su(1,1)$ SS and RIS}

In this subsection we shall construct the $su(1,1)$ SS and RIS and examine
some of their nonclassical properties.  We note that eigenstates of
complex combinations of the generators of $SU(1,1)$ are discussed in
similar ways in the recent papers \ci{T96a,B97} under the names algebraic
CS \ci{T96a} and algebra eigenstates \ci{B97}.

The three generators of $SU(1,1)$ satisfy the relations

\be\lb{19}     
[K_1,K_2]=-{\rm i}K_3,\,\,[K_2,K_3]={\rm i}K_1,\,\,[K_3,K_1]={\rm i}K_2.
\ee
To construct $su(1,1)$ SS according to (\ref{2}) we have
to solve the eigenvalue problem for one operator family

\bear\lb{20}      
A(u,v,w) = \b_iK_i = uK_- + vK_+ +wK_3 ,\\
u+v=\b_1, \,\, {\rm i}(v-u)=\b_2,\,\, w=\b_3.\nonumber
\eear
We choose the parameters $u,\,v,\,w$ and write the eigenvalue equation

\be\lb{21}  
(uK_- + vK_+ +wK_3)|z,u,v,w;k\ra = z|z,u,v,w;k\ra,
\ee
where $k$ is the Bargman index.  Consider first the series $D^{(+)}(k)$,
$k=1/2,1,...$.  The bosonic realization of $D^{(+)}(k)$ are of current
interest in quantum optics \ci{BM96}.  For example in terms of
two boson lowering and raising operators $a$ and $b$ one has

\be\lb{22}     
 K_- = ab,\,\,\,K_+ = a^\dg b^\dg,\,\,\,K_3 = (a^\dg a + b^\dg b
+ 1)/2.
\ee
This representation is irreducible in the subspaces with
 fixed eigenvalue $n_a-n_b$ of $a^\dg a -b^\dg b$, the Bargman index being
$k=(|n_a-n_b|+1)/2$.  $K_2$-$K_3$ IS in this representation can be
generated and used to improve the accuracy in the interferometric
measurements \ci{BM96}.

To solve eq. (\ref{21}) it is suitable to use the
representation of Barut and Girardello CS (BG representation) \ci{BG}.
In BG representation  $K_{\pm}$ and $K_3$ are differential operators:

\be\lb{23}   
K_+ = \eta,\,\quad K_- =
2k\,{\rm d}/{\rm d}\eta+\eta\, {\rm d}^2/{\rm d}\eta^2,\,\quad K_3 =
k+\eta\,{\rm d}/{\rm d}\eta,
\ee
where $\eta$ is a complex variable. Eq. (\ref{21}) becomes a second
order differential equation, which is easily reduced to the Kummer
equation \ci{AS}.  The solutions ($u\neq 0$) are found in the form
\ci{T96a}

\be\lb{24}   
\Phi_z(\eta;u,v,w) = N(z,u,v,w)\exp\l(c(u,w,l)\eta\r)M(k+z/l,2k,l\eta/u)
\ee
where $N(z,u,v,w)$ is a normalization constant, $M(a,b,\eta)$ is the Kummer
function  ($M(a,b,\eta) = \,_1F_1(a;b;\eta)$) \ci{AS,BE} and

\be\lb{25}   
c(u,w,l) = -\frac{1}{2u}(w+ l), \quad l = \sqrt{w^2-4uv}.
\ee
Note, $l^2=(A,A)$, where $(,)$ is the Killing
form on the Lie algebra (here $su(1,1)$) \ci{BR}.  $\Phi_z(\eta;u,v,w)$
represents normalized states $|z;u,v,w;k\ra$ if

\be\lb{26}   
|w-l| < 2|u|, \quad {\rm or}\quad  |w+l| < 2|u|.
\ee
When these normalizability conditions are broken down the functions
$\Phi_z(\eta;u,v,w)$ are still solutions of (\ref{21}), but represent
nonnormalizable states.  $|z;u,v,w;k\ra$ can  easily be expressed as series
in terms of orthonormalized eigenstates $|n+k,k\ra$ of $K_3$ using the
expansion of $M(a,b,\eta)$ in terms of powers of $\eta$. The
orthonormalized eigenstates $|k,k+m\ra$ of $K_3$ are represented by
monomials $\eta^m\,[\Gamma(2k)/(m!\Gamma(m+2k))]^{1/2}$.

\begin{eqnarray}\lb{26a}   
|z,u,v,w;k\ra &=& {\cal N}\,\sum_m^\infty g_m(z,u,v,w,k)|k,k+m\ra,\\
g_m &=&
c^m\sqrt{\frac{(2k)_m}{m!}}\,_2F_1(a,-m;2k;\zeta),\quad \zeta =
-\frac{l}{uc} = \frac{2l}{w+l}, \\
{\cal N}^{-2} &=& (1+s)^{-2k+a+a^*}\l|(1+s-s\zeta)^{-a}\r|^2\,
_2F_1\l(a,a^*;2k; \frac{-s|\zeta|^2}{\l|1+s-s\zeta\r|^2}\r)\quad\nonumber\\
              &=& {\cal N}^{-2}(s,\zeta,a), \\
s &=& -c^*c = \frac{|w+l|^2}{4|u|^2},\quad \quad
a =  k + z/l,\,\, l = \sqrt{w^2-4uv},\nonumber
\end{eqnarray}
where $_2F_1(a,b;c;z)$ is Gauss hypergeometric function \ci{BE}.  Closed
expressions of the normalization constants of  $|z,u,v,w;k\ra$, eq.
(\ref{24}), were obtained (in different parameters) by Brif \ci{B97}.

The above solution can be applied to the cases of $u=0$ or $l=0$ under
the appropriate limits $u\rightarrow 0$, $l\rightarrow 0$  in it (there is
also no problem to consider these cases separately). In other parameters
solutions of eq.  (\ref{21}) were obtained in \ci{B97} using an analytic
representation in the unit disk.

The family of $|z,u,v,w;k\ra$ contains several known types of states. The
BG CS are recovered at $v=0=w$ and the $SU(1,1)$ generalized IS (the
$K_1$-$K_2$ Schr\"odinger IS) of ref. \ci{T94} are reproduced at $w=0$ (the
identification $u = (\lam+1)/2,\,v=(\lam-1)/2$).  All the $SU(1,1)$ group
related CS with symmetry are naturally included in $|z,u,v,w;k\ra$.  It is
curious that $SU(1,1)$ CS with maximal symmetry $|\zeta;k\ra$ are
contained also in the subfamily of $K_1$-$K_2$ IS $|z,u,v;k\ra\!\equiv\!
|z,u,v,w\!=\!0;k\ra$:\, $|z\!=\!-k\sqrt{-uv},u,v;k\ra = |\zeta\! =
\!${\sm$\sqrt{-v/u}$}$;k\ra$ \ci{T94}.

$|z,u,v,w;k\ra$ are eigenstates of linear combination of three observables
$K_1,\,K_2,\,K_3$. According to the discussion in the subsection 2.3 (and
the results of \ci{T97}) these states should minimize the Robertson
inequality for the {\it three} observables if and only if they are
eigenstates of real combination of $K_1,K_2,K_3$, i.e. when $uK_- + vK_+
+wK_3$ is Hermitian. This holds when $v=u^*$ and $w$ is real, so the
states $|z,u,u^*,w;\ra$ with real $w$ are $su(1,1)$ RIS. One can check
that $|z,u,v,w;k\ra$ and the RIS $|z,u,u^*,w;\ra$ as well obey the
definition (\ref{3}) of ideal SS for the three operators $K_j$.

All $K_i$-$K_j$ Schr\"odinger IS are naturally contained in
$|z,u,v,w;k\ra$. For example $K_2$-$K_3$ IS are obtained at $v=-u^*$ and
$K_1$-$K_2$ IS -- at $w=0$.  The set of IS $|z,u,v;k\ra,\,\,v\neq 0$ is a
set of ideal $K_1$-$K_2$ SS. It is an example of a set of ideal SS which
are of the form (\ref{2}) but cannot be represented in the form (\ref{10})
of an orbit of the group Aut$(su^C(1,1)$,  i.e. in this example the second
(the orbit) construction of generalized SS is not equivalent the first and
the third ones (see proof in Appendix B). The first (the eigenstate)
and the third (the intelligent) constructions of generalized SS are
equivalent in this case since $(uK_+ + vK_+)|\psi\ra = z|\psi\ra$ is
necessary and sufficient for a state to minimize Schr\"odinger inequality
for $K_{1},\,K_2$.

In the case of one mode bosonic representation with $k=${\sm$1/4,\,3/4$},
when

\be\lb{27}    
K_- = \frac{1}{2}a^2,\quad K_+ = \frac{1}{2}a^{\dagger
2},\quad K_3 = \frac{1}{2}(a^\dagger a +\frac{1}{2}),
\ee
it is suitable to use the canonical CS representation in which $a =
{\rm d}/{\rm d}\alpha,\,\, \,a^\dagger = \alpha$.  Here we have two
independent solutions of eq.  (\ref{21}), represented by the even and odd
analytic functions $\Phi_z^{\pm}(\alf;u,v,w)$. The {\it even solution} is
given by the formula (\ref{24}) with the replacements $\eta \rar
\alf^2/2,\,\, k\rar 1/4$. Its expansion in terms of the eigenstates of
$K_3=(a^\dg a/2 + 1/4)$ (the Fock states) is given by formula (\ref{26a})
with $k=1/4$ [The BG representation (of states and operators) can be
safely used in the $su(1,1)$ irreducible subspaces, such as the subspaces
of even and odd states].  The {\it odd solution} takes the form

\be\lb{28}   
\Phi_z^{-}(\alpha;z,u,v,w) = \alpha
N_{-}\,\exp\l(c(u,w,l)\alpha^2/2\r)\, M\l(a_{-},3/2,l\alpha^2/2u\r),
\ee
with parameters
$$c = -\frac{1}{2u}(w+l),\,\, a_- = \frac{1}{4}(3 +
2z/\sqrt{-uv^\prime}),\,\,v^\prime = -\frac{1}{4u}l^2,\,\,
l = \sqrt{w^2-4uv}.$$

In this $su(1,1)$ representation the case $w=0$ was solved in
\ci{T95} where the two independent solutions $|z,u,v;\pm\ra$ were called
{\it generalized even and odd CS}.
In fact the even states $|z,u,v;+\ra$ were constructed earlier in ref.
\ci{T94}: in the BG representation $\Phi_{z^\pr}(z)$ of states
$|z^\pr,u,v;k\ra$ \ci{T94} one has to put $k=1/4$ and $z=\alf^2/2$ in order
to get the canonical CS representation of $|z^\pr,u,v;+\ra$. Since
$|z,u,v;\pm\ra$ minimize the Schr\"odinger inequality for the quadratures
of $a^2$ they are also called Schr\"odinger squared amplitude (even/odd) IS
or squared amplitude (even/odd) SS \ci{T96b}.
Other particular cases of eq. (\ref{21}) for $k=1/4,\,3/4$  were
considered in \ci{Nagel} and in the second and fourth papers of \ci{Vac}.
The general case was solved by Brif \ci{B96}.

The normalizability conditions on $u,v,w$ are the same as in the case of
$D^{(+)}(k)$.  In the alternative case of $u=0$ the solutions can be
obtained in a similar manner or by taking the appropriate limit in
(\ref{24}) and (\ref{28}) \ci{T96a,B96}. The states $|z,u,v,w;\pm\ra$
pertain the property of $|z,u,v,w;k\ra$ (noted above) to contain all
$SU(1,1)$ group related CS with symmetry.  The squeezed vacuum states
coincides with the $SU(1,1)$ CS with maximal symmetry.  It is worth to
note the following double intelligence property of the squeezed vacuum
states:  these and only these states minimize the Schr\"odinger relation
for both $q,\,p$ and $K_1,\,K_2$ pairs of observables. The squeezed one
photon states are $K_1$-$K_2$ IS only.  These properties can easily be
derived from the discussion on minimization of Schr\"odinger inequality.
The states $|z,u,v,w;\pm\ra$ minimize the Robertson inequality for the
{\it three} operators $K_1=(a^2+a^{\dg 2})/4$,\, $K_2=-i(a^2-a^{\dg 2})/4$
and $K_3=(2a^{\dg}a + 1)/4$ when $w=w^*,\, v=u^*$.

Let us examine for squeezing the constructed $su(1,1)$ SS. According to
the general prescription (see eq. (\ref{3})) the states $|z,u,v,w;k\ra
\equiv |z,\vec{\b};k\ra$ would exhibit arbitrarily strong squeezing of the
variance of the generators $K_j$ when all but $\b_j$ are let to tend to
$0$. The calculations confirm this property \ci{T96a,T96b}, i.e.,
$|z,u,v,w;k\ra$ are ideal SS.  Here the limits $\b_1\!
=\!0,\,\b_2\!=\!0$ can be taken: at $\b_1 \!=\!0,\,\b_2\!=\!0$ one gets
eigenstates of $K_3$, in which the variance of $K_3$ vanishes (is
absolutely squeezed).  The limits $\b_3\!=\!0,\, \b_1\!=\!0$ (i.e.,
$w\!=\!0,\,v\!=\!-u$) or $\b_3\!=\!0,\,\b_2\!=\!0$ (i.e.,
$w\!=\!0,\,v\!=\!u$) however can not be taken - they would violate the
normalizability constrains (\ref{26}).  For the sake of simplicity we
shall examine in greater detail the case $w\!=\!0$. The three second
moments of $K_1,\,K_2$ in $|z,u,v;k\ra$ (for $k=1/4,\,3/4$ and
$k=1/2,1,\ldots$) read \ci{T94,T97}

\be\lb{29}   
\Dlt^2 K_1 = \frac{1}{2}\frac{|u-v|^2}{|u|^2-|v|^2}\la K_3\ra,\quad
\Dlt^2 K_2 = \frac{1}{2}\frac{|u+v|^2}{|u|^2-|v|^2}\la K_3\ra,\quad
\Dlt K_1K_2 = \frac{{\rm Im}(u^*v)}{|u|^2-|v|^2}\la K_3\ra.
\ee
These variances can easily be casted in the general matrix form (\ref{8})
for $n=2$.
The mean $\la K_3\ra$ of $K_3$ in $|z,u,v,w;k\ra$ can be
calculated (for any $k$, $k=1/4,3/4,1/2,1,\ldots$, using the explicit form
of the normalization factor ${\cal N}(s,\zeta,a)$, eq. (30), according to
the following relation,

\be\lb{29a}  
\la K_3\ra
= k + s{\cal N}^2\frac{\partial {\cal N}^{-2}}{\partial s},\quad
s = -|c(u,w,l)|^2.
\ee
For $k=1/4,3/4$ the generators $K_1=(a^2+a^{\dg 2})/4$,\,
$K_2=-i(a^2-a^{\dg 2})/4$ appear as quadratures of $a^2$ and thus here
$K_{1,2}$ squeezing coincides with the "squared amplitude" squeezing
(Hillery et. al. \ci{Vac}).  The mean photon number $\la a^\dg a\ra = 2\la
K_3\ra - 1/4$.  The $K_2$ squeezing is illustrated (see also
\ci{T96a,T96b}) in Fig. 1a on the example of generalized even CS
$|z,u,v;+\ra$ ($k=1/4$, $w=0$) with $z\!=\!1,\, u\! =
${\sm$\sqrt{1+x^2}$}$,\,v \!=\!  -x,\,\,x>0$.  For convenience we take the
quadratures of $a^2$ as $\td{K}_1\!=\!
2${\sm$\sqrt{2}$}$K_1,\quad\td{K}_2 \!=\!  2${\sm$\sqrt{2}$}$K_2$, i.e.
$a^2 = (\td{K}_1 + {\rm i}\td{K}_2)/\sqrt{2}$.  Then in the ground state
$|0\ra$ the variances of the quadratures $\td{K}_1,\,\td{K}_2$ are both
equal to $1$.  According to (\ref{5}) with (\ref{Dlt_0}) a state
$|\psi\ra$ exhibits squared amplitude squeezing if $\Dlt \td{K}_1(\psi)$
or $\Dlt\td{K}_2(\psi)$ is less than $1$. Quadratic squeezing is found
also in $|z,\sqrt{1\!+\!v^2},v;+\ra$ for $z=\pm1/2,\pm1,\pm5/2$ and
negative $v$ \ci{T96a,T96b}. As Fig. 1a shows the generalized even CS
$|1,${\sm$\sqrt{1+x^2}$}$,-x;+\ra$ are $\td{K}_2$ squeezed when $x > 1.8$.
States with strong quadratic squeezing have not been pointed out so far
(the recent papers \ci{BM97,Mar}, where quadratic squeezing is also found,
appeared after \ci{T96a,T96b}).  We note that states $|z,u,v;\pm\ra$ which
exhibit quadratic squeezing are not $SU(1,1)$ group related CS  (even more
- they are not of the form (\ref{10})).

The states $|z,u,v;\pm\ra$ can also exhibit strong (but not
arbitrarily strong) {\it ordinary} (or linear) squeezing \ci{T96b}. The
quadratures $q,\,p$ are squeezed if their squared variance is less than
{\sm$1/2$}.  In $|z,u,v;\pm\ra$ we have

\be\lb{30}  
\Dlt^2 q = \frac{1}{2} + \la a^\dagger a\ra +
2\frac{{\rm Re}[(u-v)z^*]}{|u|^2-|v|^2},\quad
\Dlt^2 p = \frac{1}{2} + \la a^\dagger a\ra - 2\frac{{\rm
Re}[(u-v)z^*]}{|u|^2-|v|^2}.
\ee
On Fig. 1a we show the plot of $\Dlt^2 p(x)$ for the same states
$|1,${\sm$\sqrt{1+x^2}$}$,-x;+\ra$, $x>0$. $p$ squeezing occurs in the
interval $0\leq x \leq 3.8$. For larger $|z|$ the $p$ squeezing is
stronger and occurs in wider interval of $x$. It is worth to underline that
in the interval $1.8\leq x\leq 3.8$ the states
$|1,${\sm$\sqrt{1+x^2}$}$,-x;+\ra$ exhibit $p$ and $\td{K}_2$ squeezing
simultaneously ({\it joint squeezing} of $p$ and $\td{K}_2$).  Joint
$q$-$\td{K}_2$ squeezing occurs in $|z,\sqrt{1+v^2},v;+\ra$ with
negative $z$ and $v$ \ci{T96b} (e.g., for $z=-1$ and $-3.8\leq v\leq
-1.8$).

Quadratic and linear squeezing occurs also in SS $|z,u,v,w;\pm\ra$ with
$w\neq 0$.  For example light squeezing of $\td{K}_1$ and $\td{K}_2$ is
found in the squeezed even Schr\"odinger cats $|\alf_+\ra$ \ci{DMM} [which
are of the form (\ref{21}) for the three observables $K_1,\,K_2,\,K_3$]

\be\lb{zzeta}  
S(\zeta)|\alpha_+\ra = S(\zeta)|z;+\ra \equiv |z,\zeta;+\ra,
\ee
where $z\!=\!\alpha^2/2$ and $K_-|z;\pm\ra = z|z;\pm\ra$, $K_- \!=\!
a^2/2$.  Note that $|z,\zeta;\pm\ra = |z,u,v,w;\pm\ra$,  where

$$u\!=\!\cosh^2r,\quad v\!=\!\sinh^2r\,{\rm e}^{2{\rm i}\theta},\quad
w\!=\!\sinh(2r){\rm e}^{{\rm i}\theta},\quad
\zeta\!=\!r\,{\rm e}^{{\rm i}\theta}.$$

In $|z,\zeta;+\ra$ one has the variances

\bear\lb{32}   
\Dlt^2\td{q}_{\pm} = 1/2
+\la a^\dagger a\ra\pm {\rm Re}\la a^2\ra,\nonumber\\
\Dlt^2\tilde{X}_{\pm} = 1+2\la a^\dagger a\ra+\la a^{\dagger 2}a^2\ra \pm
{\rm Re}\la a^4\ra - \la\tilde{X}_{\pm}\ra^2, \eear
where
$\td{q}_{+}\!=\!q,\,\,\td{q}_{-}\!=\!p,\quad\td{X}_{+}\! =
\!\td{K}_1,\,\, \td{X}_{-}\!=\!\td{K}_2$. The expressions for the means
involved are written down in the paper \ci{T96a}.

On Fig. 1b plots of $2\Dlt^2q(d)$ and $\Dlt^2\td{K}_1(d)$ are shown
for the states $|z,\zeta;+\ra$ with $z\!=\!-d\!=\!-|z|,\,\,
\zeta\!=\!0.31$. In the latter the variance of $\td{K}_1$ is lightly
squeezed for $0.1<d<0.31$, and the variance of $q$ is squeezed for
$0.17\!<\!d\!<\!0.51$.  In the interval $0.17<d<0.31$ both variances are
squeezed (joint $q$-$\td{K}_1$ squeezing). The possibility for joint
squeezing of the noncommuting observables $q,\,\td{K}_1$ (or
$p,\,\td{K}_2$) is explained \ci{T96b} by the fact, that in any even or
odd quantum state the Schr\"odinger inequality reads {\sm $\Dlt^2
q\Dlt^2 \td{K}_1\geq 0$} ({\sm $\Dlt^2 p\Dlt^2 \td{K}_2\geq 0$}).

The $su(1,1)$ SS $|z,u,v,w;\pm\ra$ can exhibit other nonclassical
properties as well, in particular sub- and super-Poissonian photon
statistics \ci{T96b}. In Fig. 2b the super-Poissonian photon number
distribution in the $p$ and $\td{K}_2$ SS $|z\!=\!1,u\!\! =
\!${\sm$\sqrt{10}$}$,v\!=\!-3,w\!=\!0;+\ra$) is plotted.  Sub-Poissonian
statistics occurs in many of the states $|z,u,v;\pm\ra$, e.g.  in
$|z,u,v;+\ra$ with
$z\!=\!-0.5-5{\rm i},\,v\!=\!-0.5,\,u\!=\!${\sm$\sqrt{1.25}$} (see Fig.
2a) and $z\!=\!\pm 2.5,\, u\!=\!${\sm$\sqrt{1+x^2}$}$,\, v\!=\!x$, where
$0\!<\!x\!<\!0.5$ \ci{T96b}. In both cases of the non-Poissonian
statistics the photon distributions exhibit well pronounced oscillations.
We note that the above pointed nonclassical states with sub-Poissonian
statistics are neither $q$ or $p$ nor $\td{K}_1$ or $\td{K}_2$ squeezed.

Squeezing and statistical properties of other subsets of algebraic CS
$|z,u,v,w;k\!=\!\frac 14,\frac 34\ra$ are studied in the recent papers
\ci{T96a,T96b,BM97,Mar} and in the second paper of \ci{Vac}. In \ci{Mar}
it was wrongly concluded that all $|z,u,v;+\ra$ with real $u,\, v$ are
super-Poissonian.
\vspace{6mm}

\bt{ll}
\hspace{-1.5cm}\input{fg1a.pic} &  \input{fg1b.pic}\\
                                   &
\et
\begin{center}
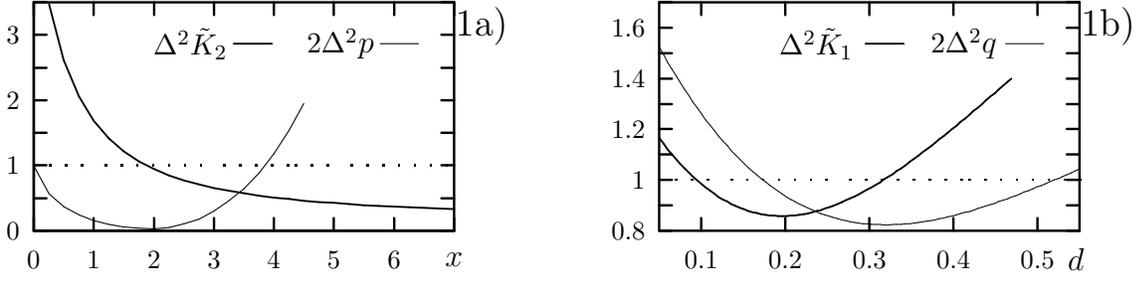
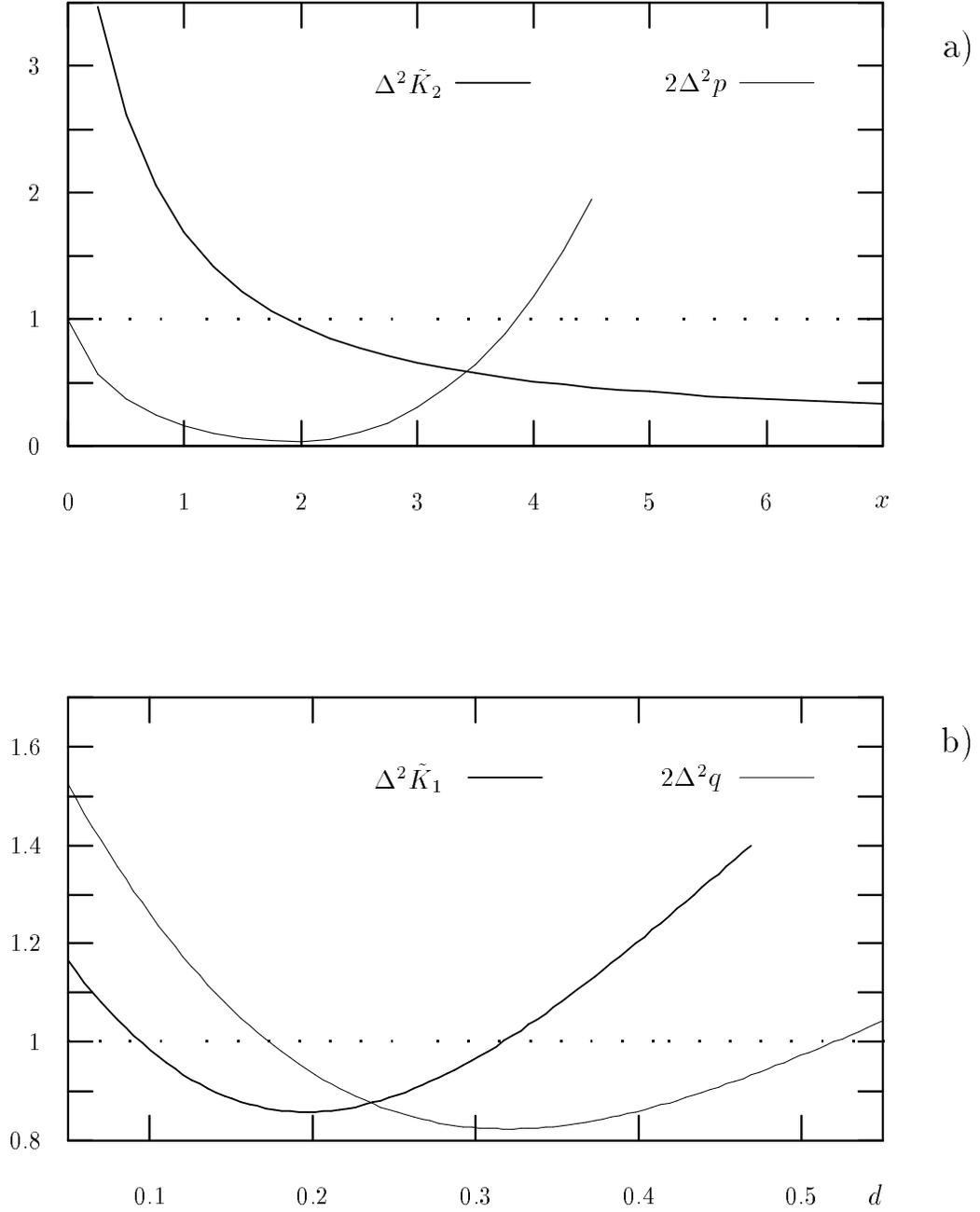
{\small Fig.1. Squeezing of quadratures of $a$ and $a^2$ in
the $su(1,1)$ even SS $|z,u,v,w;+\rangle$.}\\[1mm]
{\small a) Variances of $p=i(a^\dagger-a)/\sqrt{2}$ and $\tilde{K}_2
= i(a^{\dagger 2}-a^2)/${\small$\sqrt{2}$} in generalized even CS $\qquad$\\
$\qquad |1,${\small$\sqrt{1+x^2}$}$,-x,0;+\rangle$,
 $x>0$. Joint $p$, $\tilde{K}_2$
squeezing occurs in  $1.8\!<x\!<3.8$. $\qquad$\\[1mm]
b) Variances of $q=(a^\dagger+a)/${\small$\sqrt{2}$} and
$\tilde{K}_1=(a^{\dagger
2}+a^2)/${\small$\sqrt{2}$} in ordinary squeezed even CS\\[-1mm]
 $|z,\zeta;+\rangle$, eq. (\ref{zzeta}), for $\zeta\!=\!0.31$,\,\,
$z\!=\!-d$,\,\, $d\!>\!0$.
Joint $q$, $\tilde{K}_1$ squeezing in $0.17<\!d<\!0.31$.}
\end{center}\vspace{3mm}

\begin{center}
\begin{tabular}{cc}
\hspace{-10mm}\input{fg2a.pic}& \input{fg2b.pic}$\quad$\\[-2mm]
                                  &
\end{tabular}
\begin{tabular}{ll}
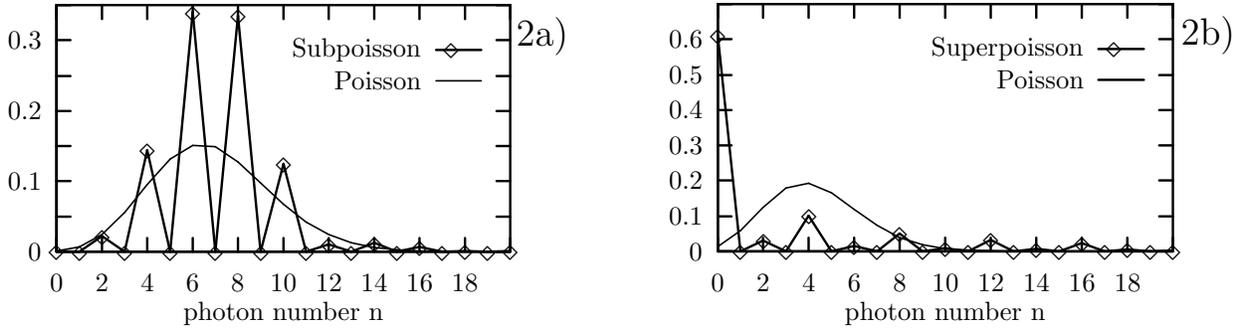
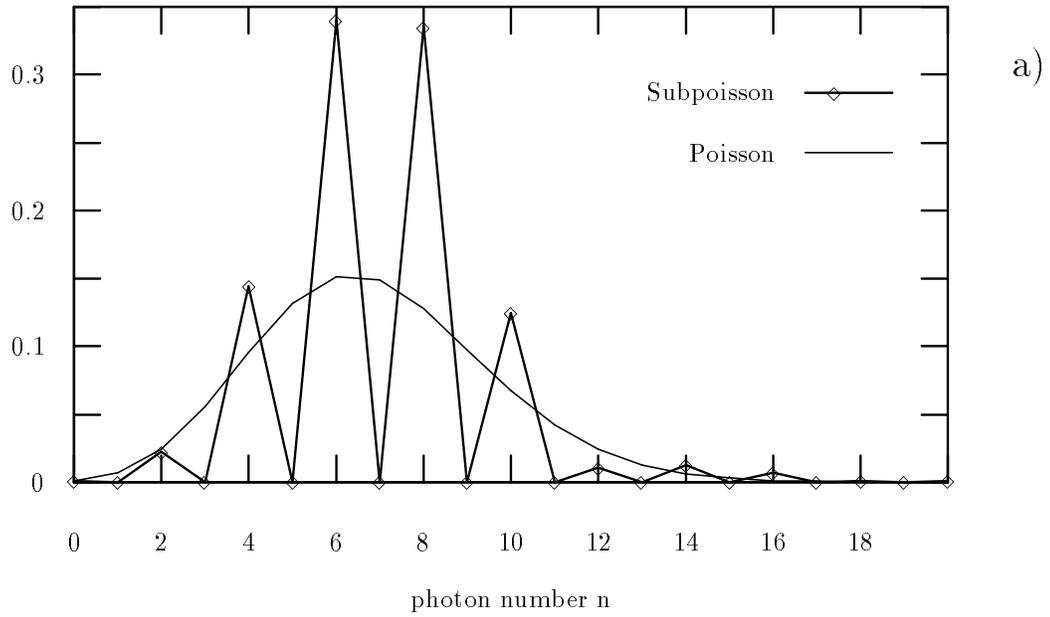
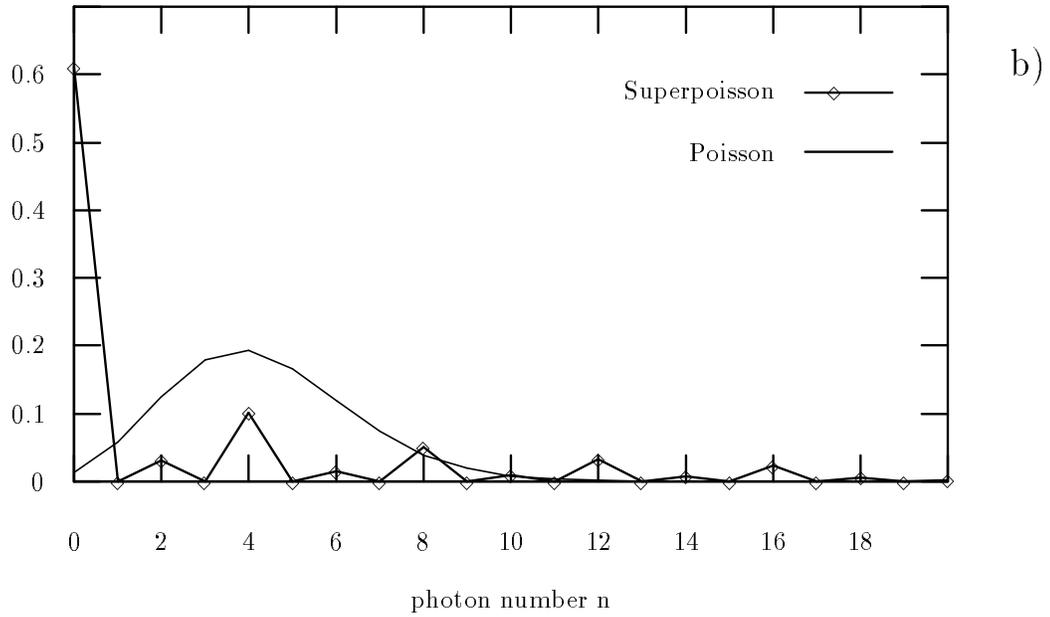
{\small Fig. 2. Non-Poissonian photon distributions in generalized even
CS $|z,u,v;+\rangle$:} & \\
{\small a) $|$-$0.5$-$5i$,$\sqrt{1.25}$,-$0.5$;+$\rangle$, $Q<0$
$(Q=-0.21)$, $\langle a^{\dagger}a\rangle=7.06$}; & \\
{\small b) $|$1,$\sqrt{10}$,-$3$;+$\rangle$, $Q>0$, $\langle
a^{\dagger}a\rangle=6.88$ (this is $p$, $\tilde{K}_2$ joint squeezed state
as in Fig. 1a).} & \\[-1pt]
{\small Poissonian distributions with mean photon numbers $7.06$
and $6.88$ are also shown}.&
\end{tabular}
\end{center}\vspace{3mm}

For physical realization of a new set of states it is important to know
the general form of Hamiltonian operator which preserves the set stable in
the time evolution. It was shown \ci{T94} that the $su(1,1)$ states
$|z,u,v;k\ra$, which minimize the Schr\"odinger uncertainty relation for
$K_{1,2}$, are stable under the action of $H \sim K_3$ only. For the case
of $k=1/4,3/4$ it is the free field Hamiltonian only which keeps the whole
set of $|z,u,v;k\ra$ stable \ci{T96b}. However this Hamiltonian can not change
the value of $|v|$. Therefore there is no unitary evolution process in
which the squared amplitude SS $|z,u,v;k\ra$ can be generated from BG CS
$|z;k\ra$ or from eigenstates of $a^2$ (in particular from Glauber CS
$|\alf\ra$).

For a large subset of $|z,u,v,w;k\ra$ there is an other possibility for
generation from states which are  finite superpositions of orthonormalized
eigenstates $|n+k,k\ra$ (in particular of Fock states). Indeed,
$|z,u,v,w;k\ra$ can be represented as $S(\zeta)|\psi_0(u,v,w)\ra$, where in
accordance with the discussion after eq. (\ref{10}) the "reference" vector
$|\psi_0\ra$ is {\it not independent} of the parameters $u,v,w$ (see
\ci{T96a}).  When in (\ref{24}) $a=-n$ (i.e., the quantization condition
$z\!=\!-(k+n)${\sm$\sqrt{l^2}$}\,$\equiv z_n$ is imposed) the "reference"
state $|\psi_0(u,v,w)\ra$ is a finite superposition of $|n+k,k\ra$ which
for $k\!=\!${\sm$1/4,\,3/4$} are number states.  Finite superposition of
number states in principle can be experimentally constructed \ci{Jan}.
Then the one mode $su(1,1)$ SS which are squeezed finite superpositions of
number states can be generated using the latter as input in the degenerate
amplifier scheme.  The $su(1,1)$ SS in the form of eq.  (\ref{zzeta}) can
be generated in the same scheme, using as input the ordinary even/odd CS.

A scheme for generation of {\it two mode} $su(1,1)$ SS in the form of
$K_2$-$K_3$ IS was presented recently by Luis and Perina \ci{BM96}.
A generation scheme for several subsets of {\it one mode} $su(1,1)$
algebraic CS, in particular, for subsets of Schr\"odinger $K_1$-$K_2$ and
$K_2$-$K_3$ IS is presented very recently \ci{BM97}. In both schemes a
photon number measurement in the process of generation is involved, i.e.
the total evolution process is not unitary, which is in accordance of the
stable evolution results \ci{T94,T97,T96b}.

In the scheme of Brif and Mann \ci{BM97} two light beams of modes $a$ and
$b$ are mixed in the non-degenerate parametric amplifier, the mode $a$
being beforehand squeezed in the degenerate parametric amplifier. These
processes are described by interaction Hamiltonians

\be\lb{H12}   
H_1 = \frac{1}{2}(g_1a^{\dg 2} + g_1^*a^2),\quad
H_2 = \frac{1}{2}(g_2a^{\dg}b^{\dg} + g_2^*ab),
\ee
where parameters $g_{1,2}$ depend on the pump and the media nonlinear
characteristics. In \ci{BM97} the input state was taken as a two mode vacuum,
$|\psi_{\rm in}\ra  = |0\ra_a|0\ra_b$. If  a measurement of $a$-mode photon
number is performed in the output beam (with the result $n$) and an
$SU(1,1)$ transformation $\exp({\rm i}\omega K_2)\exp({\rm i}\vphi K_3)$
is performed on the mode $b$, then for $\omega,\,\vphi$ suitably chosen,
the final state $|\psi_{\rm fin}\ra \equiv |\lam,n\ra$ is an eigenstate of
$\lam K_1 -{\rm i}K_2$:

\be\lb{psi_f}   
(\lam K_1 -{\rm i}K_2)|\psi_{\rm fin}\ra= (k+[n])\sqrt{1-\lam^2}|\psi_{\rm
fin}\ra,
\ee
where $[n]$ is the integer part of $n$, $k=1/4\,(3/4)$ for
even (odd) $n$, $\lam = 1/\cosh\omega = \sqrt{|\chi|^2-1}/|\chi|$,\,
$|\chi| > 1$ ($ 0< \lam < 1$). The parameter $\chi$ is related to $g_{1,2}
= |g_{1,2}| {\rm e}^{{\rm i}\t_{1,2}}$ and the interaction times $t_{1,2}$
according to $\chi = (\tanh|g_1t_1|/\sinh^2|g_2t_2|)\exp[i(\t_1-2\t_2)]$
\ci{BM97}. The choice of the $SU(1,1)$ parameters is $\vphi = \arg\chi$ and
$\tanh\omega = 1/|\chi|$.

We see that for a given $\lam$ and $n$ this scheme produces only one
eigenstate $|\lam,n\ra$ of the operator $\lam K_1 -{\rm i}K_2 = \frac
12[(\lam +1)K_- + (\lam-1)K_+]$. In particular, the even joint squeezed
states $|z,u,v;+\ra,\,\,z=1,\,v=-x,\,\,u=\sqrt{1+x^2}$, the statistical
properties of which are presented on Figures 1a and 2b, are beyond the
family of final states $|\lam,n\ra$.  To produce the whole family
$|z,\lam;k\ra$, $z\in C$, of the eigenstates of the operator
$\lam K_1 -{\rm i}K_2$ one has to modify the Brif and Mann scheme in order
to introduce extra free parameters. A suitable modification is that which
includes two displacements $D(\g_1)$ and $D(\g_2)$ on the mode $b$ prior
to and after the above described $SU(1,1)$ transformation. Then   an input
state of the form $|\alf\ra_a|0\ra = D(\alf) |0\ra_a|0\ra_b$, under a
certain relation among $\g_{1,2},\,\vphi$ and $\alf$, would be finally
transformed into $|\psi_{\rm fin}\ra$,

\be\lb{z,eta,n}    
(\lam K_1 - {\rm i}K_2)|\psi_{\rm fin}\ra= \mbox{\sm
$\frac{1}{2}$}[\mbox{\sm $\frac{1}{2}$} + n + \zeta(\g_1,\omega,\vphi)]
\sqrt{1-\lam^2}|\psi_{\rm fin}\ra,
\ee
where
\begin{eqnarray}\lb{gamma} 
\zeta = |\g_1|^2 + \g_1^2\coth\o\, {\rm e}^{{\rm i}\vphi} -
\g_1\td{\alf} + \mbox{\sm $\frac 12$}[\g_2^2\coth\frac{\omega}{2} -
\g_2^{*2}\tanh\frac{\omega}{2}],\\
\g_2 =
\tanh\frac{\omega}{2}\l[\g_1\sinh\frac{\omega}{2}\, {\rm e}^{{\rm
i}\vphi/2} - (\g_1^* + 2\g_1\chi - \td{\alf})\cosh\frac{\omega}{2}\,
{\rm e}^{-{\rm i}\vphi/2}\r],\\ \td{\alf} =
(\cosh\frac{\o}{2}+\sinh\frac{\o}{2}) {\rm e}^{{\rm i}\vphi/2}\,
{\rm Re}\td{\b} - {\rm i}(\cosh\frac{\o}{2}-\sinh\frac{\o}{2})
{\rm e}^{{\rm i}\vphi/2}\, {\rm Im}\td{\b}\quad\\ \td{\b} =
(\g_1+2\g_1^*\chi^*)\cosh\frac{\o}{2}\, {\rm e}^{{\rm i}\vphi/2} + (\g_1^*
+ 2\g_1\chi)\sinh\frac{\o}{2}\, {\rm e}^{-{\rm i}\vphi/2} + \qquad\nonumber\\
 \g_1\sinh\frac{\o}{2}\,\tanh\frac{\o}{2}\, {\rm e}^{{\rm i}\vphi/2} -
\g_1^*\cosh\frac{\o}{2}\, \coth\frac{\o}{2}\, {\rm e}^{-{\rm i}\vphi/2}.
\end{eqnarray}
This $|\psi_{\rm fin}\ra$ is of the form of $|z,u,v;k\ra$,
$|u|^2-|v|^2=1$, with $u=(\lam+1)/2\sqrt{\lam}$, $v =
(\lam-1)/2\sqrt{\lam}$ and $z = (1/2)(1/2 + n + \zeta)
\sqrt{(1-\lam^2)/\lam}$. The $p$-$\td{K}_2$ joint SS shown on Figure 2b
is produced by this generation scheme when $|\chi| = \coth\o = 1.0004$,
$\vphi = 0$, $n=0$ and $\g_1 = 0.91$. The sub-Poissonian even SS shown on
Figure 2a is produced when $|\chi| = \coth\o = 1.08$, $\vphi = 2.374$,
$n=0$ and $\g_1 = 1.01 + {\rm i}0.36$.


\section{Conclusion}

We have examined three possible ways of generalization of the one mode
canonical squeezed states (CSS) to the case of $n$ arbitrary observables
$X_j$ on the basis of the Robertson uncertainty relation and the group of
automorphisms of the corresponding complexified Lie algebra. The
eigenstate way of generalization appeared as the most general one.  The
cases of $N$ pairs of canonical observables $p_\nu,\,\,q_\nu$ (which span
the algebra $h_N$ and generates the group $H_N$)  and the three quasi-spin
observables $K_j$ (the generators of $SU(1,1)$) are considered as
examples. The case of nilpotent algebra $h_N$ is most symmetric in the
sense that the three constructions of $\vec{X}$-SS by means of eqs. (2),
(10) and (13) are equivalent to each other. For the simple $su(1,1)$
algebra such equivalence is lacking.  In this context it is shown that the
overcomlete family of eigenstates $|z;k\ra$ of $su(1,1)$ lowering operator
$K_-$ (the Barut-Girardello coherent states (CS)) and the continuous
family of ideal SS  $|z,u,v;k\ra$ (eigenstates of $uK_- + vK_+$) are not
orbits of the group Aut$(su^C(1,1)) \supset SU(1,1)$. Such continuous sets
are examples of proper algebraic coherent states \ci{T96a}.

Eigenstates $|z,u,v,w;k\ra$ of general complex combinations of $K_j$ are
explicitly constructed using the analytic Barut-Girardello CS
representation. These are ideal SS for the three $SU(1,1)$ generators.
Their nonclassical properties are analyzed in the case of $k=1/4,3/4$,
when $K_j$ are quadratic combinations of boson/photon operators $a$ and
$a^\dg$.  Intelligent even states $|z,u,v,0;+\ra$  are pointed out which
exhibit sub-Poissonian photon statistics and joint linear and quadratic
amplitude squeezing. These sub-Poissonian and joint SS could be produced
from the canonical CS by an appropriate modification of the recently
proposed Brif and Mann generation scheme \ci{BM97}.  Joint SS of the field
could be useful in improvement of the sensitivity of the interferometric
measurements.
\vspace{3mm}

{\bf Acknowledgments}. The author thanks prof. V.I. Man'ko for valuable
comments. The work is partially supported by the Bulgarian Science
Foundation under Contracts No. F-559, F-644.
\newpage

\section{Appendix}
{\bf A. Equivalence of the three definitions of SS for $2N$
canonical observables}

  The three generalized definitions of SS for $h_N$ algebra elements
$p_\nu,\,\,q_\nu$ are given by the eqs.  (\ref{2}), (\ref{10}) and
(\ref{13}) with $X_j = Q_j$, $Q_\nu = p_\nu$ and $Q_{N+\nu} = q_\nu$. In
ref. \ci{T97} it was proven that in case of $X_j = Q_j$ the equality
(\ref{13}) entails (\ref{2}) with $n_{\rm c} = N$ and (\ref{10}) as well. What
remains to be proven in order to establish the equivalence of the three
definitions (\ref{2}), (\ref{10}) and (\ref{13}) is that eq. (\ref{2})
with any $n_{\rm c}\geq 1$ also entails  (\ref{13}).

{\large \it Proposition 1.} {\it If a state $|\psi\ra$ is an eigenstate of
an operator $A(\vec{\b},\vec{\g}) = \b_\nu p_\nu
+\g_\nu q_\nu$,

\be\lb{33}   
A(\vec{\b},\vec{\g})|\psi\ra = z|\psi\ra,
 \ee
then $|\psi\ra$ is an eigenstate of $N$ new boson operators $A_\nu$, which
are linear in $p_\nu$ and $q_\nu$}.

{\it Proof}. Let $|\psi\ra$ satisfy eq. (\ref{33}).  In coordinate
representation the solution to (\ref{33}) is given by an exponent of a
quadratic (see eq. (\ref{17b})),

\be\lb{34}      
\psi_z(\vec{q},\vec{\b},\vec{\g}) = \tilde{\cal
N}\exp\l[-\vec{q}{\cal M}\vec{q} + \vec{\cal N}\vec{q}\r], \ee
where

\be\lb{35}  
{\cal M}_{\nu\mu} = -({\rm i}/2){\b_\nu}^{-1} \g_\mu, \,\,\, \vec{\cal
N}_\nu = {\rm i}{\b_\nu}^{-1}z,
\ee
The matrix ${\cal M}$ is symmetric and  the state
$\psi_z(\vec{q},\vec{\b},\vec{\g})$ is normalizable iff
${\cal M}^* + {\cal M}$ is positive definite.

Let us now treat ${\cal M}$ and $\vec{\cal N}$ in eq. (\ref{35}) as
given, $z$ as arbitrary and consider (\ref{35}) as algebraic equations for
$\g$ and $\b$. The solution is easily seen to be not unique. For an
arbitrary $\vec{\b} = (\b_1,\b_2,\ldots,\b_N)$ the vector $\vec{\g}$ is
uniquely determined as $\vec{\g} = -2{\rm i}{\cal M}\vec{\b}$. Thus for
every state $\psi(\vec{q},{\cal M},\vec{\cal N})$ of the form (\ref{34}),
i.e. for
every given ${\cal M}$ and $\vec{\cal N}$,  we have an $N$ complex
parameter family of linear in $p_\nu,\,\,q_\nu$ operators $A({\cal
M},\vec{\cal N};\vec{\b})$ which have the same state $\psi(\vec{q},{\cal
M},\vec{\cal N})$ as their common eigenstate. Since parameters $\b_\nu$
are free we may
choose $N$ vectors $\vec{\b}_{(\mu)}$ (i.e. a matrix $\b$), consider
then $N$ operators from this family, %

\be\lb{36}    
A_\mu({\cal M},\vec{\cal N};\b) = \b_{\mu\nu}p_\nu +
\g_{\mu\sig}(\b)q_\sig,\quad
\g_{\mu\sig}(\b) = -2{\rm i}(\b {\cal M}^{\rm T})_{\mu\sig},
\ee
and try to subject $A_\mu({\cal M},\vec{\cal N};\b)$ to the canonical
boson commutation relations (\ref{18}). The latter require $\b\g^\dg -
\g\b^\dg = i$ and $\b\g^{\rm T} - \g\b^{\rm T} = 0$. Substituting here
$\g = -2i\b {\cal M}^{\rm T}$ we get equations for $\b$, %

\be\lb{37}     
\b ({\cal M}^* + {\cal M}^{\rm T})\b^\dg = \frac{1}{2},\quad
\b {\cal M} \b^{\rm T} - \b {\cal M}^{\rm T}\b^{\rm T} = 0.
\ee
The matrix ${\cal M}$ is symmetric, thereby the second equality in
(\ref{37}) is satisfied identically. The real and symmetric matrix
${\cal M}^* + {\cal M}^{\rm T} = {\cal M}^* + {\cal M}$ is positive
definite and therefore is congruent to a multiple of unity by the matrix
$\b$ which is symplectic or is a product of one orthogonal and one
diagonal positive matrix \ci{Gant}.  Thus eq. (\ref{37}) always has a
solution (not unique) for $\b$ which ensures the canonical boson relations
(\ref{18}) for $A_\mu({\cal M}, \vec{\cal N};\b)$. End of proof.
\vspace{3mm}

\bc {\bf B. The sets of $K_1$-$K_2$ SS $|z,u,v;k\ra$ and the
Barut-Girardello CS are not orbits of Aut$(su^C(1,1))$}
\ec

Unlike the case of $h_N$ (canonical observables) the eigenstates of linear
combinations of $su(1,1)$ operators $K_j$ (quasi-spin observables) and of
other semisimple Lie algebras can not always be presented in the form
(\ref{10}), i.e. as orbit of unitary operators from Aut$(su^c(1,1))$. In
subsection 2.2 we have noted that any Aut$(L^C)\equiv G_{\rm A}$ group related
CS with symmetry is of the form of (\ref{2}), i.e.  is an eigenstate of a
complex linear combination of algebra operators $X_i$, while the inverse
is generally not true.  In this Appendix we provide such a "negative"
examples of generalized SS: we shall prove that the $su(1,1)$
Schr\"odinger intelligent states (IS) $|z,u,v;k\ra$ (constructed first in
\ci{T94}) can't be represented in the form of eq. (\ref{10})).  These states
can exhibit arbitrary strong squeezing in $SU(1,1)$ generators $K_1$ or
$K_2$ when $v \rar \pm u$ ($|v|<|u|$), i.e.  they are ideal SS of
the form (\ref{2}).  But the family of states $|z,u,v;k\ra$ is not neither
$SU(1,1)$ nor Aut$(su^C(1,1))$ unitary orbit of any fixed reference vector
$|\psi_0\ra$, as we are going to prove bellow.

The $K_1$-$K_2$ SS $|z,u,v;k\ra$ are defined as eigenstates of $A_-(u,v) =
uK_- + vK_+$,

\be\lb{38}
A_-(u,v)|z,u,v;k\ra = z|z,u,v;k\ra, \ee
and for $k=1/2,1,\ldots$ and $k=1/4,3/4$ are explicitly constructed in
subsection 3.2.

{\large \it Proposition 2.} {\it There is no Hilbert space vector
$|\psi_0\ra$ such that the family $\{|z,u,v;k\ra\}$ be an Aut$(su^C(1,1))$
unitary orbit of $|\psi_0\ra$, i.e.}

\be\lb{39} |z,u,v;k\ra =
U(u,v,z)|\psi_0\ra,\quad U \in {\rm Aut}(su^C(1,1)), \quad U^\dg
=U^{-1}.\ee

{\it Proof}. Let us suppose the inverse, i.e. let (\ref{39}) holds
for some state vector  $|\psi_0\ra$, which is independent of the
parameters $u,v,z$. We shall show that this leads to a contradiction (and
therefore (\ref{39}) is impossible).

In view of $U\in {\rm Aut}(su^C(1,1))$ the transformation $K_i \rar
U^\dg(z,u,v)K_iU(z,u,v)\equiv K^\pr_i$ is linear in $K_j$. One has
\be\lb{40}
U^\dg(z,u,v)A_-(u,v)U(z,u,v) = \mu(z,u,v)K_- + \nu(z,u,v)K_+
+\sig(z,u,v)K_3, \ee
Eqs. (\ref{40}), (\ref{39})  and (\ref{38}) imply that
\be\lb{41}
(\mu(z,u,v)K_- + \nu(z,u,v)K_+ +\sig(z,u,v)K_3)|\psi_0\ra = z|\psi_0\ra
\ee
It is worth noting that the invariance of the Killing form $B(X,Y)$ for
$su^C(1,1)$ \ci{BR},
\be\lb{42}
 \sig^2(z,u,v) - 4\mu(z,u,v)\,\nu(z,u,v) = 4uv = B(A_-,A_-), \ee
requires that neither $\sig$ and $\mu$ nor $\sig$ and $\nu$ can vanish
simultaneously. At $v=0$ (then $A_-(u,0) = uK_-$ and we put $u=1$)
(\ref{42}) reads $\sig^2(z) - \mu(z)\nu(z) = 0$. It is interesting to note
that using only the invariance of the Killing form one  can easily derive
that the orthonormalized eigenstates $|k,k+n\ra$ of $K_3$ can't satisfy
eq.  (\ref{41}), i.e.  $|\psi_0\ra \neq |k,k+n\ra$, $n=0,1,\ldots$ [For
$k=1/4,3/4$ $|k,k+n\ra$ coincide with the Fock states $|n\ra$]. For the
sake of brevity henceforth we write $|n\ra$ instead of $|k,k+n\ra$.

Generally $|\psi_0\ra$ is a superposition of $|n\ra$,
\be\lb{43}
|\psi_0\ra =  \sum_{n=0}C_n|n\ra,\quad \sum_n|C_n|^2 = 1. \ee
Substituting (\ref{43}) into (\ref{41}) we obtain the
recurrence relations for the coefficients $C_n$,
\be\lb{44}
\mu(z,u,v)\sqrt{n+1}C_{n+1} + \nu(z,u,v)\sqrt{n}C_{n-1}
+\sig(z,u,v)(n+k)C_n = zC_n. \ee
It is sufficient to prove that (\ref{39}) is impossible for some
subset of states $|z,u,v;k\ra$. We shall carry out the proof for the
subsets $\{|0,u,v;k\ra\}$ and $\{|z,1,0;k\ra\}$. Note that the states
$|z,1,0;k\ra$ are the BG CS $|z\ra$, $|z,1,0;k\ra = |z;k\ra$. Let us first
choose the subset $|0,u,v;k\ra$. We easily see that if $C_0 = 0 = C_1$
then all $C_n = 0$.  Moreover both $C_0$ and $C_1$ are nonvanishing. For
$n=0$ and $n=1$ (\ref{44}) produces
\be\lb{45}
C_1 \mu = k\sig C_0, \quad \mu C_2\sqrt{2} = -\nu C_0 - (k+1)\sig C_1.
\ee
From $C_1 \mu = k\sig C_0$ we derive that $C_1 =0 \lrar C_0=0$. Indeed,
if e.g. $C_0 =0$ but $C_1 \neq 0$, then $\mu =0$ and the second equation
in (\ref{45}) yields $\sig =0$ which contradicts to (\ref{42}).
Thus $\,\,\, C_0\neq 0 \neq C_1$. \,\,
%
%
The two eqs. (\ref{45}) tell us also that the ratios $\mu/\sig$ and
$\nu/\sig$ must be $u$ and $v$ independent.  Denoting $ \mu/\sig =
kC_0/C_1 \equiv a_1,\quad \nu/\sig = a_2$ we rewrite (\ref{42}) as
$\sig^2(0,u,v)(1-a_1a_2) = 4uv$, which at $v=0$ produces $\sig(0,1,0) = 0$
[since $1-a_1a_2 =0$ would lead to $0 = 4uv$].  On the other hand is eq.
(\ref{40}), which now reads $U^\dg(0,u,v)A_(u,v)U(0,u,v) =
\sig(0,u,v)[a_1K_- + a_2K_+ + K_3]$. Herefrom at $v=0$ we obtain
$U^\dg(0,1,0)K_-U(0,1,0) = 0$ which is impossible for the unitary operator
$U\neq 0$ [$U(0,1,0) = 0$ would lead to $|k,k\ra = 0$ since $|0\ra \equiv
|k,k\ra = U(0,1,0)|\psi_0\ra$]. This contradiction proves that the
continuous set of states $|0,u,v;k\ra$ (which are annihilated by $uK_- +
vK_+$) is not an Aut$(su^C(1,1))$ orbit of any reference vector.
End of proof.

{\large \it Proposition 3.} {\it The set of Barut-Girardello CS $|z;k\ra$
is not an {\rm Aut}$(su^C(1,1))$ unitary orbit}.

{\it Proof}. The BG CS constitute a subset of $|z,u,v;k\ra$, $|z;k\ra =
|z,1,0;k\ra$. We follow the scheme of the proof of Proposition 2. Let us
first note that if the unitary operator $U(z) = U(z,1,0)$ (obeying
(\ref{39}) with $u=1$, $v=0$) exists, it cannot commute with $K_-$,
otherwise $K_-|z\ra = U(z)K_-|0\ra =0$. The recurrence relations for the
coefficients $C_n$ in (\ref{43}) are given by eq. (\ref{44}) and this time
the Killing form vanishes identically with respect to $z$:

\be\lb{46}
\sig^2(z) - 4\mu(z)\nu(z) = 0.
\ee
Again $C_0 \neq 0 \neq C_1$, and instead of (\ref{45}) now we have

\be\lb{47}
C_1 \mu = (z-k\sig)C_0, \quad \mu C_2\sqrt{2} = -\nu C_1 - [(k+1)\sig -z]
C_0.
\ee
After some consideration we get from (\ref{47}) that the ratios
$\sig/\mu\equiv a_2 \neq 0,\,\nu/\mu\equiv a_1\neq 0$ and $z/\mu\equiv
a_3$ are independent of $z$. Then we rewrite (\ref{40}) in the form

\be\lb{48}   
U^\dg(z)K_-U(z) = za_3[K_- + a_1K_+ + a_2K_3],
\ee
which at
$z=0$ yields the contradiction $U^\dg(0)K_-U(0) = 0$. For $z\neq 0$
another contradictions arise: Let us apply both sides of  (\ref{48}) to
$U^\dg(z)|0\ra$. This gives $0 = [K_- + a_1K_+ + a_2K_3]U^\dg(z)|0\ra$.
In view of $U(z)\in {\rm Aut}(su^C(1,1))$ we have $U(z)[K_- + a_1K_+ +
a_2K_3]U^\dg(z) = \mu^\pr(z)K_- + \nu^\pr(z)K_+ + \sig^\pr(z)K_3
\neq 0$ and thus $0 = [\mu^\pr(z)K_- + \nu^\pr(z)K_+ +
\sig^\pr(z)K_3]|0\ra = [\nu^\pr(z)K_+ + \sig^\pr(z)k]|0\ra$. One sees that
the last equality is possible if and only if $\nu^\pr = 0 = \sig^\pr$.
Then, we have $(1-\mu^\pr(z))U(z)K_-U^\dg(z) = U(z)[a_1K_+ +
a_2K_3]U^\dg(z)$. Applying this to $U(z)|0\ra$ (and noting that
$\mu^\pr\neq 1$, otherwise the commutator $[K_-,U(z)]$ vanishes) we have
$0 = [a_1K_+ +a_2k]|0\ra$, which is impossible since $a_1 \neq 0 \neq
a_2$.  End of the proof.

Remark: The Propositions 2 and 3 are valid for any Hermitian representation of
$su(1,1)$ for which the BG CS $|z\ra$ and the eigenstates $|z,u,v\ra$ of
$uK_- + vK_+$  exist. These results can be extended to semisimple Lie
groups.

\end{document}